\documentstyle[12pt]{article}
\begin{document}

 \newcommand{\um}[1]{\"{#1}}
 \newcommand{\uck}[1]{\o}
 \renewcommand{\Im}{{\protect\rm Im}}
 \renewcommand{\Re}{{\protect\rm Re}}
 \newcommand{\ket}[1]{\mbox{$|#1\protect\rangle$}}
 \newcommand{\bra}[1]{\mbox{$\protect\langle#1|$}}
 \newcommand{\proj}[1]{\mbox{$\ket{#1}\bra{#1}$}}
 \newcommand{\expect}[1]{\mbox{$\protect\langle #1 \protect\rangle$}}
 \newcommand{\inner}[2]{\mbox{$\protect\langle #1 | #2 
\protect\rangle$}}

\begin{center}
{\Large  \bf Can a falling tree make a noise in two forests at the 
same time?} \\
\vspace{0.2 in} 
Aephraim M. Steinberg \\
Department of Physics \\
University of Toronto \\
Toronto, ON M5S 1A7 \\
Canada \\
(version of \today)
\end{center}

\vspace{0.5 in}
\begin{quote}
{\bf Abstract}\\
It is a commonplace to claim that quantum mechanics supports the old 
idea that a tree falling in a forest makes no sound unless there is
a listener present.  In fact, this conclusion is far from obvious.
Furthermore, if a tunnelling particle is observed in the barrier 
region, it collapses to a state in which it is no longer tunnelling.
Does this imply that while tunnelling, the particle can not have any 
physical effects?  I argue that this is not the case, and moreover,
speculate that it may be possible for a particle to have effects on two 
spacelike separate apparati simultaneously.
I discuss the measurable consequences of such a 
feat, and speculate about possible statistical tests which could 
distinguish this view of quantum mechanics from a ``corpuscular'' 
one.  Brief remarks are made about an experiment underway at Toronto
to investigate these issues.
\end{quote}
\vspace{0.5 in}

Jean-Pierre Vigier has long been at the center of some of the deepest
controversies in quantum mechanics, his work bearing on questions of 
determinism, locality, and the very nature of light.  At this meeting 
in his honour, we have seen that these disputes rage on, and have 
heard analyses of experiments as old as the Michelson-Morley and as 
recent as the latest work on correlated photon pairs.  I hope it is 
fitting to extend this range yet further into the future, and attempt 
to draw conclusions from an experiment so new that we have only begun 
to build it at the University of Toronto.  

It is of course well-known that many aspects of quantum mechanics, 
most notably the EPR experiments and the Aharonov-Bohm effect, are 
nonlocal; but also that causality is always enforced, {\it i.e.}, 
that no information is ever conveyed faster than light.  The
tension between these concepts 
points to a gap between our theory and our understanding of this
theory.  It seems worthwhile to study the varieties of nonlocality
predicted by quantum mechanics, so as to begin improving this state of
affairs.  In this paper, I make some brief remarks about 
some ``local''
intuitions I believe have survived quantum theory in the minds of 
many practicing physicists, and speculate about whether or not they 
are in fact consistent with our theory.  I will describe a set of 
experiments we are currently preparing at Toronto for the purpose of 
investigating these issues in the laboratory.

What do we mean when we say that quantum mechanical nonlocality does 
not violate causality?  In the most well-known example, the case of EPR 
correlations, what occurs at one point may depend nonlocally on what 
occurs at some far-separated point.  However, neither occurrence 
depends on anything we {\it  control} at a far-separated point.  In the 
language of Shimony, quantum mechanics may violate ``outcome 
independence,'' but it preserves ``parameter 
independence''\cite{Shimony=1986,Jarrett=1984}.  No 
information about our own decisions gets transmitted faster than 
light; only {\it random} information about which of several possible 
outcomes was realised is ``propagated'' (one might better say
{\it created}) in a nonlocal fashion.

The other best-known example of nonlocality in quantum mechanics, the 
Aharonov-Bohm effect, shows that the phase of a particle may be 
affected by a magnetic field the particle never actually enters.  
However, this absolute phase is unmeasurable.  Only by having a 
portion of the particle's wave function go around the other side of 
the confined field can we measure a {\it relative} phase and thus acquire 
information about the quantity of enclosed flux.  This requires us to 
spatially close the loop, removing any possibility of acquiring the 
desired information in a superluminal fashion.

Published discussions of ``nonlocality of a single 
particle''\cite{Tan=1991,Hardy=1994,Greenberger=1995} have 
so far relied on examples where there are in fact at least two 
particles at play (one typically masquerading as a ``local oscillator'').  
The fact that one cannot tell which of the two 
one observes is used as an indication that one's interpretation ought 
not  ascribe a local existence to either one of the particles.

Finally, it has been observed that in the presence of gain, loss, or
tunnelling, the peak of a transmitted wave function may appear to 
traverse a region faster than the vacuum speed of light $c$.  However,
it has also been shown that no information is conveyed by this
superluminal propagation, and that the obstacle merely serves as
a sort of analog computer, {\it predicting} within a certain level
of approximation what the future of the wave function ought to look
like\cite{Chiao=1997,Mitchell=1997}.

It has been shown in the context of relativistic quantum 
field theory \cite{Eberhard=1989}
 that information transfer faster than light is in 
general impossible.  More precisely formulated, the statement is as 
follows: no action performed at a spacetime point $({\bf 
r}_{0},t_{0})$ has any effect on the expectation value for any 
measurement performed at a point $({\bf r}_{1},t_{1})$ s.t. 
$|c(t_{1}-t_{0})|<|{\bf r}_{1}-{\bf r}_{0}|$.

Let us then consider a slightly different question.  Granted, a cause can 
have no spacelike-separated effect.  But can a cause have two 
{\it different} effects which are spacelike-separated from each
{\it other} (not, of course, from the cause itself)?  A simple example 
is a radio broadcast, where two receivers on opposite sides of the 
transmitter may simultaneously receive a message; there is clearly no 
problem with causality here.  There is also no problem with locality: 
the antenna sends out many photons, some going one way and some going 
the other.  Were it only to send out a single photon, no more than 
one of the receivers could detect it.

Or is this so?  Can a single particle/wave influence ``detectors'' at two 
spacelike-separated positions?  I believe most practicing physicists
share with the layman the 
classical intuition I shall call ``corpuscular,'' which says that
such a thing is impossible.  Even if a particle's behaviour is 
described by a wave equation, in the end all of the acceptable 
alternate outcomes we expect quantum mechanics to offer us generally 
involve a particle being in one place or another.  This is enforced by 
collapse: if I detect a particle at receiver 1, the state of the 
system collapses to one in which the wave function of the particle no 
longer extends out to receiver 2.  The implication here is that the 
wave function is ``epistemological,'' rather than ``ontological,'' to 
follow the usage of Aharonov {\it et al.}, who have recently argued 
that we should recognise the physical reality of a wave function 
itself\cite{Aharonov=1993PLA,Aharonov=1993PRL,Unruh=1994}.

The question becomes significantly less clear if we remove the 
collapse event from the equation.  A particle has an extended wave 
function.  Even if its interactions are purely local, it may 
therefore influence two separated measuring devices simultaneously.  
This is not to say that two separated detectors will at the same 
instant fire at the instigation of a single particle, but only that 
the wave functions of both detectors may be modified by the particle's 
passage.  If the interaction between the particle and either device is 
extremely strong, then the two possible final states of the device 
corresponding to presence or absence of a passing particle will be 
nearly orthogonal.  This orthogonality will lead to decoherence, or an 
effective ``collapse,'' such that one detector's firing will 
completely inhibit any effect of the particle on a spacelike separated 
detector\cite{von_Neumann=1955,von_Neumann=1983,Zurek=1991}.  

Following the ideas of Aharonov {\it et al.} on ``weak measurement,ее let
us consider instead the limit where this interaction between the particle 
and the detectors is exceedingly 
weak\cite{Aharonov=1988,Aharonov=1990,Reznik=1995}.  The predictions of quantum
mechanics are clear: both detectors' wave functions are modified by the
passing particle, but neither is modified very much.  By ``not very much,'' 
we mean that the inner product of the initial and final states is close
to unity.  Since no measurement can distinguish perfectly between 
nonorthogonal states, we can never be {\it certain} whether either 
detector
was influenced by the particle.  Therefore on no occasion will we find 
ourselves
in the uncomfortable position of having to say that an individual particle
definitely interacted with both spacelike separated detectors.

But what can we say after many particles have passed the detectors,
each one having a wave function split evenly between the two (see Fig. 
1)?  
Eventually, each detector will have accumulated a measurable shift.
We may attribute this shift to the detector having been shifted slightly
by that half of the particles which interacted with it -- or alternatively,
to its having been shifted by {\it half} of {\it each} particle, since half
of each particle's wave function interacted with it.  Is it possible to
distinguish these pictures?  Can we conclude (not on a case-by-case basis,
but for an ensemble of passing particles) that {\it both} detectors must
sometimes have been influenced by the same particle?

It seems likely that a statistical procedure will in fact make such an 
investigation possible.  Let $N$ particles each be split into a two-part 
wavefunction, interacting with $2N$ detectors $A_i$ and $B_i$ ($i$ 
running
from $1$ to $N$).  Let the initial state of the detectors be described by
gaussians centered at $0$:
\begin{equation}
\Psi(x_A) = exp(-x_A^2/4\sigma^2) \equiv G_A(0,\sigma).
\end{equation}
Further let the action of a single passing particle be to shift the
wave function of the relevant detector by an amount $\Delta \ll \sigma$
to $G_A(\Delta,\sigma)$ or $G_B(\Delta,\sigma)$,
the inequality enforcing the ``weakness'' of the measurement, {\it i.e.},
the non-orthogonality of initial and final detector states.

The expectation values of both $x_A$ and $x_B$ have shifted by 
$\Delta/2$ due to the passage of the particle, and of course, the
expectation value of $x_A-x_B$ remains unchanged at $0$.  This
does not yet resolve the question.  What must be examined is the
{\it distribution} of $x_A-x_B$.  Its initial width of $\sqrt{2}\sigma$
will be found to have grown slightly (to$\sqrt{2\sigma^2 + \Delta^2}$)
in the final state
\begin{equation}
\frac{G_A(\Delta,\sigma)G_B(0,\sigma)\ket{a} + 
G_A(0,\sigma)G_B(\Delta,\sigma)\ket{b}}{\sqrt{2}} \; .
\end{equation}
This is because the orthogonality of the free-particle states
$\ket{a}$ and $\ket{b}$ make the effective density matrix 
describing the detectors an incoherent mixture of $A$ having
shifted and $B$ having shifted.  That is, the information contained
in the particle makes it always possible in principle to go back
and ascertain which detector has shifted (albeit by a nearly
immeasurable amount).

The situation becomes significantly more subtle if this information
is erased\cite{Scully=1991,Kwiat=1992,Chapman=1995,Herzog=1995}.  
That is, suppose the particle's two paths are recombined
and the particle is postselected to be in the initial superposition
\begin{equation}
\frac{\ket{a} + \ket{b}}{\sqrt{2}} 
\end{equation}
(this is what I mean by a ``conditional measurement'').
This then leaves the detectors in the combined state
\begin{equation}
K\left\{ G_A(\Delta,\sigma)G_B(0,\sigma) + 
G_A(0,\sigma)G_B(\Delta,\sigma)\right\} \; ,
\end{equation}
where the normalisation constant $K \equiv [2 \left( 
1 + |\inner{G_A(0,\sigma)}{G_A(\Delta,\sigma)}|^2 \right)]^{-1/2}$.
We see, however, that even this entangled state implies an
anticorrelation between the two detectors, and therefore expect that the 
uncertainty in $x_A - x_B$ will still have grown.  (Finding $A$ at
$x=0$, for example, would leave $B$ centered somewhat to the right
of $\Delta/2$, while finding $A$ at $x=\Delta$ would leave $B$ further
to the left.)

It is not yet clear whether such an anticorrelation must always persist,
but there is at least one situation which seems worthy of further
investigation.  Consider a particle tunnelling from left to right across
a rectangular barrier.  At early times, prepare it in a wavepacket incident
from the left.  At late times, it will be in a superposition of a reflected
and a transmitted wavepacket; in certain regimes, the peak of the latter
will have appeared superluminally with respect to the peak of the incident
packet\cite{Hauge=1989,Landauer=1994,Steinberg=1993PRL,Enders=1993,%
Spielmann=1994,Steinberg=1995ANG} (these effects, and 
their lack of implications for causality, are
discussed in \cite{Chiao=1997}).  Now suppose that we observe the 
particle
on the right-- in effect, this constitutes a projection onto the 
superluminally
transmitted portion of the wave packet.  Following the weak measurement
formalism alluded to earlier, Fig. 2 shows the ``conditional probability
distribution'' for the particle's position as a function of time (see 
\cite{Steinberg=1995PRL,Steinberg=1995PRA}).
Note that this distribution is normalized, and reproduces exactly the
state preparation and post-selection; at intermediate times, it is
composed of a diminishing series of peaks near the entrance of the
barrier, and a growing series of peaks near the exit.  Physically, this
distribution represents the magnitude of the effect which standard
quantum mechanics predicts the particle would have on a detector at
the position and time in question.  A striking feature of the group 
delay time for quantum tunnelling is that it saturates as a finite 
value even as the barrier thickness grows\cite{Hartman=1962,%
Steinberg=19941D2D,Spielmann=1994,Landauer=1994}.  For extremely thick 
barriers, the shrinking peaks on the left and the growing peaks on the 
right will be almost entirely spacelike-separated, and yet each, when 
integrated over time, will contain nearly 100\% of the particle density
(contingent, of course, on the post-selection, whose success rate 
falls exponentially with barrier thickness).

It is not pointless to stress the procedure of a conditional measurement
and the meaning of the quantities graphed in the figure.  These 
measurements, while described by the standard apparatus of quantum
theory, are not the type of measurement we are accustomed to 
dealing with, and the expectations we have of quantum measurements
are not always fulfilled by them.  Consider an ensemble of particles
prepared in some state $\ket{i}$.  Let each particle interact with a
detector via the familiar von Neumann interaction ${\cal H}_{int}(t)
= g(t) x_p \cdot p_D$ such that subsequent to the interaction, the
mean detector position $\expect{x_D}$ will have shifted by an amount
equal (for an appropriately normalised $g(t)$)
 to the particle position $\expect{x_p}$.  (The {\it uncertainty} in the 
detector position may be small or large, corresponding,
respectively, to ideal quantum measurements or to Aharonov's ``weak''
measurements.  In the former case, an effective collapse occurs, but
in the latter case, the back-action on the particle may be arbitrarily
small.)  But before recording the detector position, we may study the
particle further.  We may, for example, test whether or not it is in
the final state $\ket{f}$.  If this test fails, we discard the
detector reading, and only when the desired final state is reached do
we record the position of the detector; this additional projection,
or post-selection, constitutes the heart of a conditional measurement.
We are now free to ask what the mean position of the detector is.
The insight of Aharonov and his coworkers was that due to the 
time-reversibility of quantum evolution (in the absence of collapse),
this quantity ought to depend as much on the post-selected state
$\ket{f}$ as on the preselected state $\ket{i}$.  In the limit of a
sufficiently weak measurement, they found that inferred value
of a measured operator $A$ is
\begin{equation}
\expect{A}_{wk} = \frac{\bra{f}A\ket{i}}{\inner{f}{i}} \; .
\end{equation}
It is easy to see that in the ``standard'' case $f=i$, this reduces to
the usual expectation value.  Now, for a weak measurement, any
individual observation will have such a large uncertainty that very
little information will be obtained; but by repeating many such
observations, all the necessary statistics can nevertheless be
built up.  One has thus sacrificed case-by-case precision in the name
of ``gentleness,'' of avoiding uncontrollable quantum 
back-action\cite{Aharonov=1988,Aharonov=1990,Reznik=1995}.

These conditional weak measurements have many interesting
linearity-related properties.  For example, if {\it either} the initial or
the final state is an eigenstate of the operator to be measured, the
result is simply the corresponding eigenvalue.  This leads to
counter-intuitive results when combined with linearity.  For example,
if one prepares a particle in an eigenstate $S_z = +1/2$ and post-selects
it in an eigenstate $S_x = +1/2$, consider a measurement of the
spin-projection operator $[S_z + S_x]/\sqrt{2}$.  It will yield a
value of $1/\sqrt{2}$, which is outside the eigenvalue spectrum of
the measured operator.  Of course, no individual event will give
definite evidence of such a value, because of the large uncertainty
of the measurement; it is only when the average of many events is
taken that this anomalously large shift will be 
observed\cite{Aharonov=1988,Aharonov=1993NKE}.

We are investigating the possibility that a similar anomaly will 
occur in connection with tunnelling\cite{Steinberg=1995PRL,Steinberg=1995PRA,%
Chiao=1997}.  Namely, a wave packet prepared
to the left of a barrier will be certain to affect a detector at
the entrance face of the barrier, as long as the detector is left
open for a time longer than the wave-packet width.  At the same time,
a wave packet {\it detected} (post-selected) on the right of the
barrier will be certain to have affected a detector at the exit
face of the barrier.  Due to the superluminality of the transmission
indicated in Fig. 2, the time windows for these two detectors may
be spacelike-separated for a thick enough barrier.  Both detectors 
have then acquired a {\it certain} shift due to the nonlocal passage of 
the tunnelling particle, to be contrasted with the {\it possible} (and
anticorrelated) shifts of Eq. (4).  It seems possible that both 
detectors will therefore shift, {\it without} any increase in the 
uncertainty of their difference.  A corpuscular model, on the other 
hand, would necessarily cause that uncertainty to grow, since only 
one detector would be affected by each particle.  We are looking into 
the possibility of deriving rigorous inequalities to describe the 
predictions of such corpuscular models.  We will also investigate the 
generalisation of Eq. (5) to higher moments, in order to compare the 
quantum-mechanical predictions for such conditional distributions with 
the requirements of corpuscularity.  Iannaccone has already begun 
studying such quantities\cite{Iannaccone=pre}.

At the same time, we are building an atom-optics experiment which will
let us directly test these questions, along with related issues to do 
with decoherence, dissipation, quantum-mechanical motion in 
time-dependent potentials, et cetera.  We will start with a 
sample of laser-cooled Rubidium atoms in a magnetic trap, and use a 
tightly focussed beam of intense light detuned far to the blue of the 
D2 line to create a dipole-force potential for the 
atoms\cite{Rolston=1992,Miller=1993FORT,Davidson=1995}.  Using a 5W 
laser at 532 nm, we will be able to make repulsive potentials with 
maxima on the order of the Doppler temperature of the Rubidium vapour.
Acousto-optical modulation of the beam will let us shape these 
potentials with nearly total freedom, such that we can have the atoms 
impinge on a thin plane of repulsive light, whose width would be on 
the order of the cold atoms' de Broglie wavelength.  This is because 
the beam may be focussed down to a spot several microns across 
(somewhat larger than the wavelength of atoms in a MOT, but of the 
order of that of atoms just below the recoil temperature, and hence
accessible by a combination of cooling and selection techniques).  
This focus may be rapidly displaced \cite{Steinberg=1996ICAP,Rudy=1997} by 
using acousto-optic modulators.  As the atomic motion is in the 
mm/sec range, the atoms respond only to the time-averaged intensity, 
which can be arranged to have a nearly arbitrary profile.  

While eventually we plan to study ultracold (ideally, Bose-condensed)
atoms in free fall, initial 
experiments will operate in the presence of the magnetic trap, relying 
on the dipole force both to perform additional velocity selection and 
to split the trap potential into an asymmetric double-well
(cf. \cite{Davis=1995}).  A ``classical'' barrier
much wider than the wavelength will be used to 
adiabatically sweep all atoms with 
energies below the barrier height off to one side; atoms with higher 
energies will return to the center of the trap.  We will thus have a 
secondary trap with velocity-selected atoms dominating the local 
density (see Fig. 3).  By subsequently reducing the barrier width to a 
minimum (and simultaneously increasing the barrier height), we will be 
able to move into the quantum-mechanical regime.  The tunnelling rate 
through this thin barrier will be monitored by imaging the trap and measuring 
the leakage from one well into the other.  (For feasible parameters,
predicted tunnelling rates are of the order of 1\% per secular period,
and the secular period will be on the order of tens of milliseconds, 
far shorter
than the lifetime of the trap.)  In the longer term, 
further cooling techniques will be used and the atoms will be allowed 
to fall freely, impinging on a barrier in otherwise free space.  
Multiple barriers, and combinations of magnetic and RF fields, will be
used to form more complicated structures, such as Fabry-Perot cavities
for the atoms.

The advantage of observing atoms tunnelling through these micron-scale 
barriers is that the atoms' internal degrees of freedom offer an ideal
way to perform the types of conditional measurements implicit in
Figure 2.  (Not only will this allow us to investigate the questions 
of locality I have raised here, but also simpler issues such as the 
prediction that the particle will almost never be seen in the center 
of the tunnel barrier, even when it is found to be transmitted.)  In 
analogy with the Larmor clock championed by 
B\"uttiker\cite{Buttiker=1985}, we will use 
the Zeeman and/or hyperfine levels of the Rubidium ground state to 
keep track of where the atoms have been.  A focussed probe beam can be 
used to optically pump the atoms between different 
long-lived ground states.  It is straightforward to calculate, and to 
measure in free
space, the time rate of change of the polarisation under the
influence of this beam.
The transmitted and reflected atoms will be spatially separated, and their 
polarisations can therefore be measured independently by standard 
optical techniques.  
In this way, we will be able to measure how long
each subensemble spent in the region illuminated by the probe beam.
If the probe beam is pulsed, we can investigate a specific area of 
time as well as space.  The analogy to a spin tunnelling through a 
magnetic-field region which causes Larmor precession is clear, 
although the use of optical beams makes it much easier to create a 
confined interaction region (with both spatial and temporal 
variability).  The existence of multiple hyperfine and Zeeman levels 
only enriches the problem, and as discussed in 
\cite{Steinberg=1995PRL}, there are aspects of weak measurements which 
only become clear for the Larmor clock when dealing with spin $> 1/2$. 

For the idealized tests we are aiming at, we will eschew optical pumping
due to the dissipation inherent in its spontaneous emission phase.
At the start, we plan to use  the entirely coherent
process of stimulated Raman scattering as a probe.
More elaborate investigations of dissipation-induced decoherence will
follow at a later date.  
Not only is the connection between
decoherence and the measurement of a particle in a forbidden region
interesting (any measurement which in fact collapses the atom into the
barrier region necessarily imparts enough energy to the atom that it is
now classically allowed, and evolves freely at subsequent times), but
tunnelling itself in the presence of dissipation has long been an important
problem\cite{Caldeira=1983,Ranfagni=1990}.  
As for the size of the probe beam, it too will be diffraction-limited
at at least several microns, so in order to study scales smaller than that of
the barrier, we will produce small displacements, carrying out a differential
measurement.

In order to address the question from the title, we will apply two 
simultaneous probe beams, and study their joint action of the atoms'
internal states.
For example, if one beam is adjusted to pump into higher Zeeman 
levels and the other to lower Zeeman levels, we can study whether half 
the atoms move in each direction or whether the two effects cancel 
out, leaving the atoms in their initial states.  Such a cancellation
would imply that an atom can be simultaneously 
influenced by two separate beams while it is tunnelling.  (In practice,
we will be dealing with length scales on the order of microns, and
will of course not be able to check locality directly.  We will have 
probe pulses separated sufficiently so that to cross both of them 
would require velocities larger than any characteristic speeds of the 
atoms in the experiment, and test whether the atoms nevertheless 
interact simultaneously with both beams; this is analogous to 
performing a test of Bell's Inequalities without having fast enough 
timing to close the loophole.)  We expect to see that this is in fact 
the case: that while both probes get shifted by the particles, the 
difference of the two shifts does not become any more uncertain.  
Coupled with appropriate inequalities, such an observation would serve
as an experimental demonstration that in quantum mechanics, not only 
can an atom falling through a forbidden barrier make a sound, but it 
can make sounds in two places at the same time-- but {\it only} if no
one is there to listen too closely.

{\bf Figure Captions}

\begin{itemize}

\item Two spacelike-separated receivers may be simultaneously affected
by a classical radio broadcast.  What if the receivers are instead 
looking at different arms of an interferometer, both of which are
simultaneously traversed by {\it each} particle which enters the 
device?  As discussed in the text, an ideal detection event will lead
to collapse (inhibiting interference, of course), but ``weak'' 
measurements will create a situation in which both detectors $D_{1}$ 
and $D_{2}$ pick up small shifts, only measurable after many many 
trials.  The question of corpuscularity can be addressed by asking 
whether the {\it difference} $D_{1}-D_{2}$ increases in uncertainty 
during this process or whether the two detectors shift in parallel 
with no extra dispersion.  We are looking for quantitative 
inequalities for testing the hypothesis of corpuscularity.  While the 
experiment drawn in this figure is not expected to disprove the 
hypothesis, we speculate in the text that a modified version 
involving tunnelling particles may well offer certain surprises.

\item For a tunnel barrier extending from $x=-5$ to $x=+5$ (not shown),
this sequence of snapshots shows the evolution of a tunnelling 
probability distribution.  At early times, this is equivalent to
$|\Psi(x)|^{2}$ for a wave packet incident from the left, while at 
late times it mimics a wave packet exiting (superluminally, for 
appropriate parameters) on the right.  The formalism for calculating 
these conditional probability distributions is explained in 
\cite{Steinberg=1995PRL,Steinberg=1995PRA}.  Note that at no time is 
there significant probability for the tunnelling particle to ``be''
near the center of the barrier, and that (to within the packet width)
the distribution ``jumps'' instantaneously from the entrance to the
exit of the barrier.  This behavior persists for arbitrarily wide
barriers.

\item This is a schematic of our proposed experimental setup.  Atoms 
initially cooled to sub-Doppler temperatures and trapped in a 
quadrupole magnetic trap will be subjected to the dipole force of an 
intense, focussed blue-detuned laser.  If the beam waist is far 
greater than the atoms' de Broglie wavelengths, the motion is 
essentially classical: all atoms with high enough energies surmount 
the barrier, and all others are reflected.  As we slide the barrier 
through the atom cloud at velocities small with respect to the atomic 
motion, the lowest-energy atoms are adiabatically carried away from 
the trap center.  By abruptly decreasing the beam waist and 
increasing the intensity, we will now be able to study the tunnelling 
of this velocity-selected sample through a quantum-mechanical barrier.
(In experimental practice, the beam waist will be as small as possible
from the start, but rapid dithering of the beam position will be used 
to create a time-averaged potential with an effective waist several 
times larger.)

\end{itemize}

\vspace{0.1in}
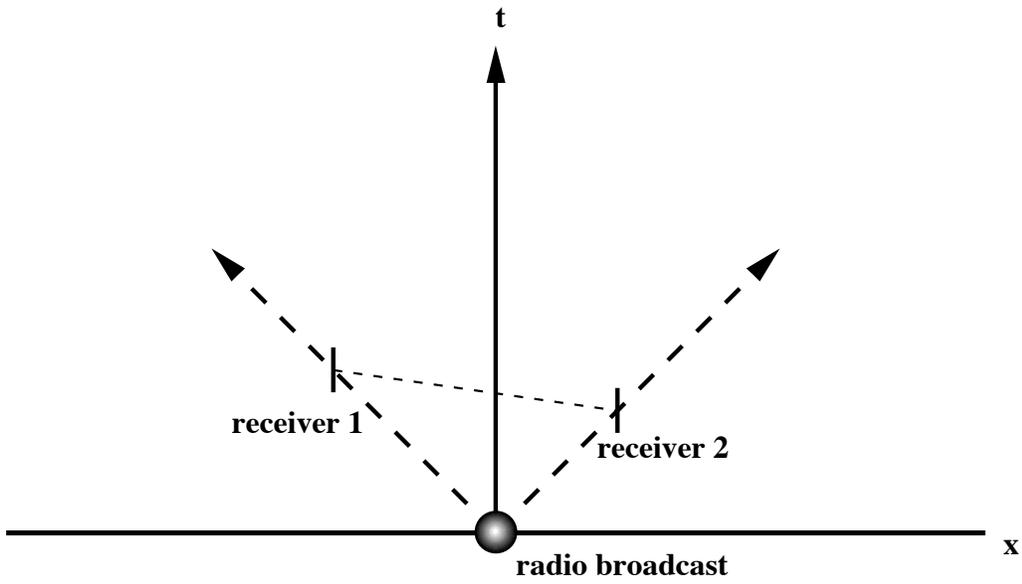
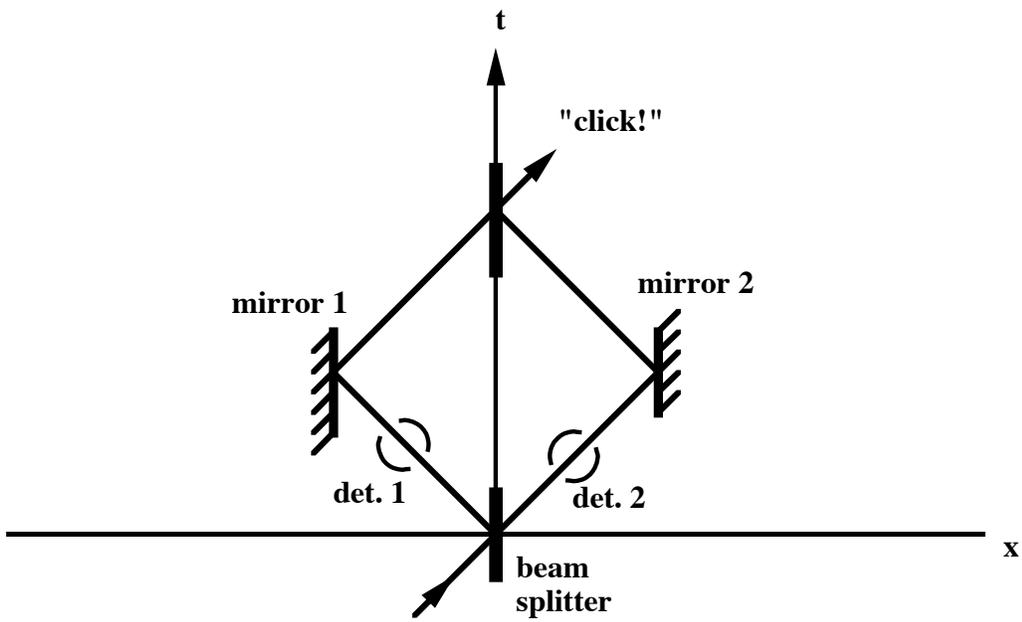
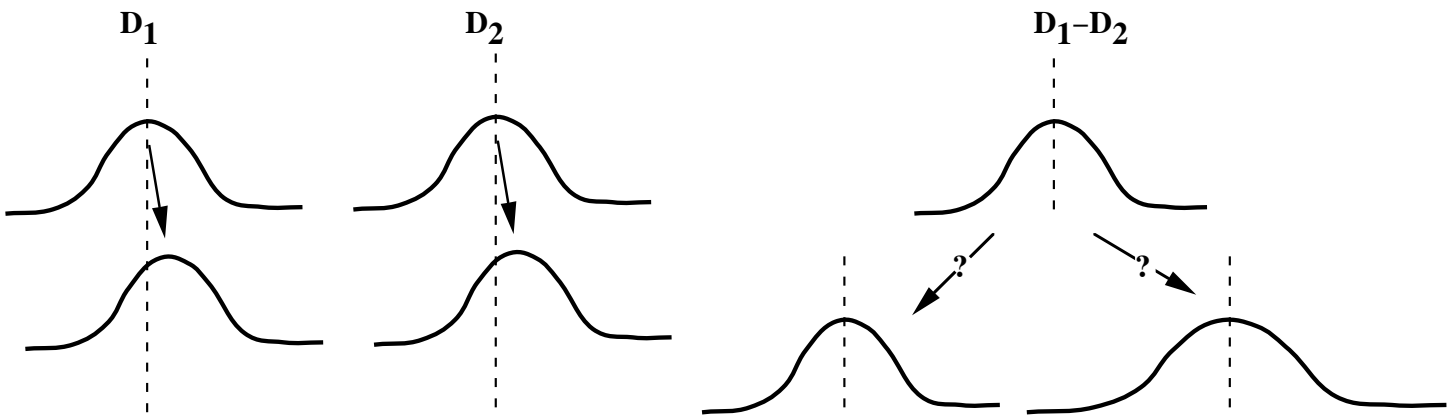
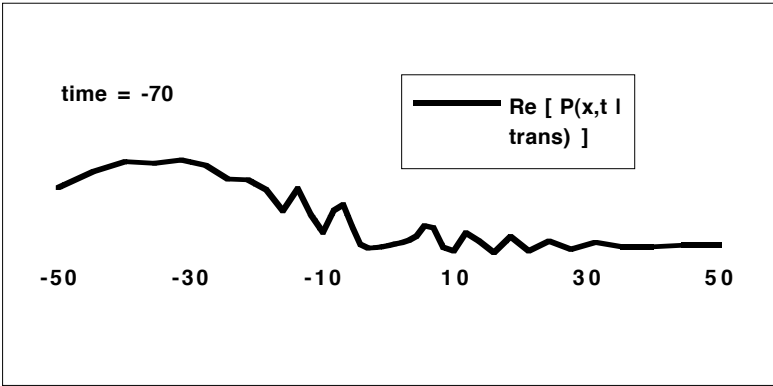
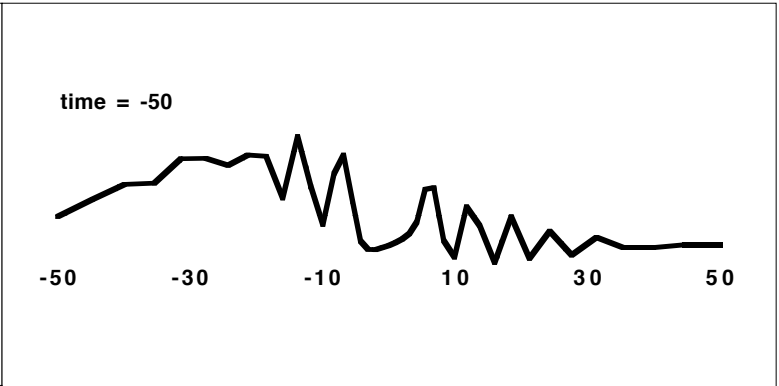
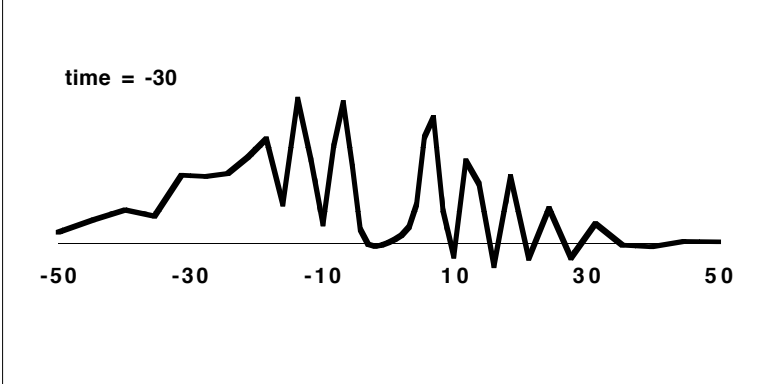
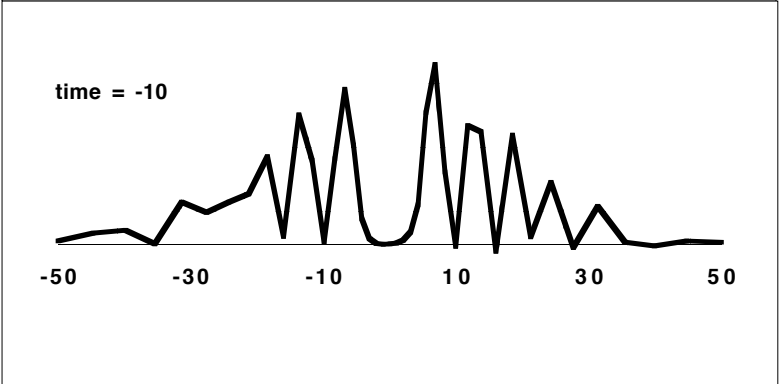
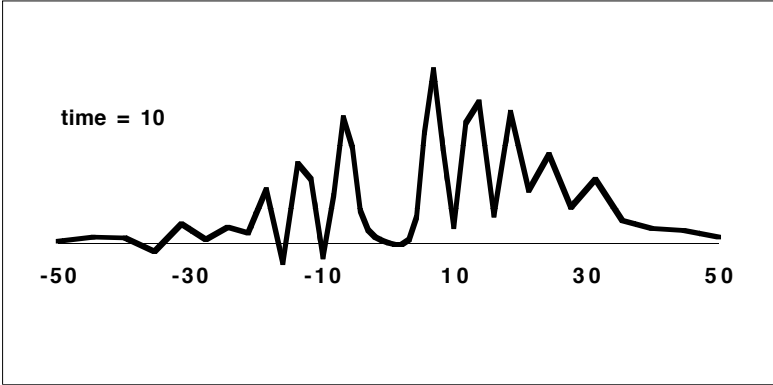
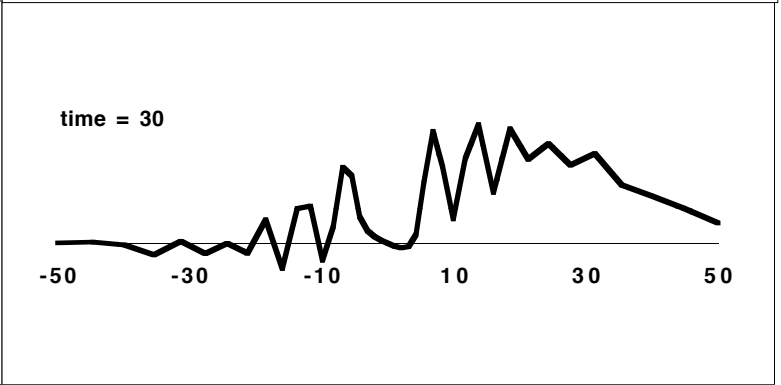
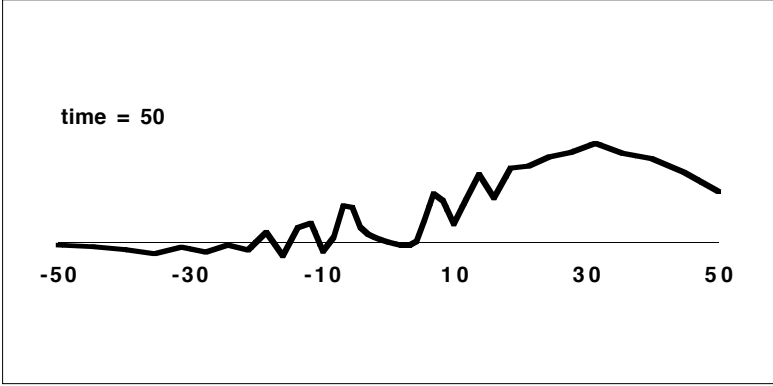
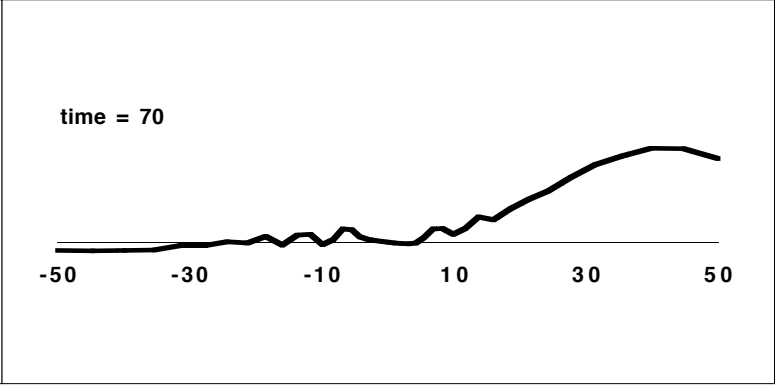
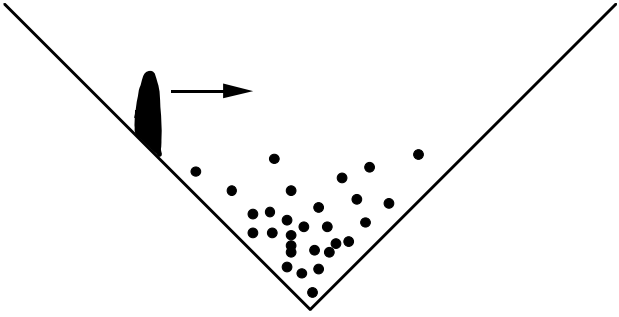
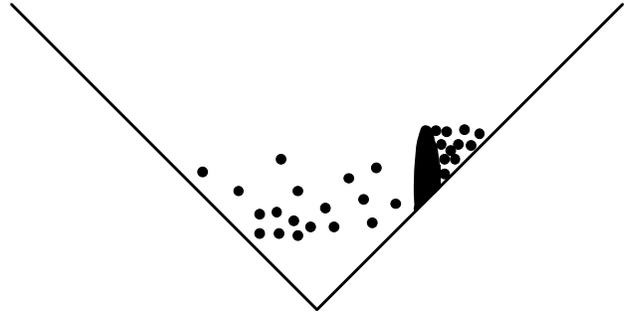
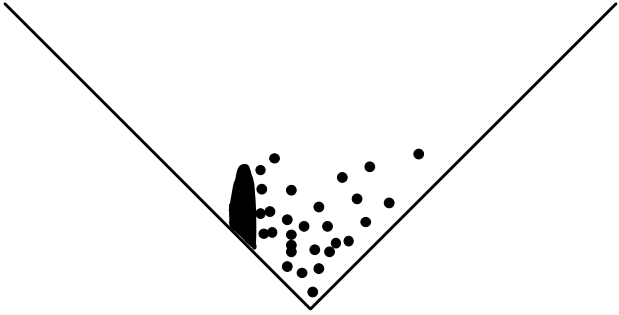
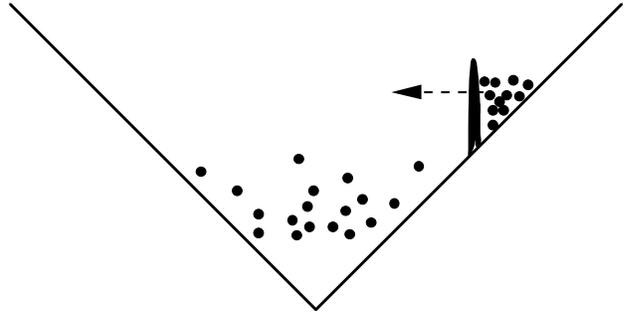
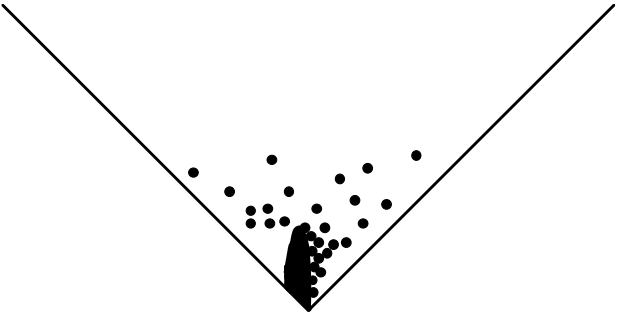
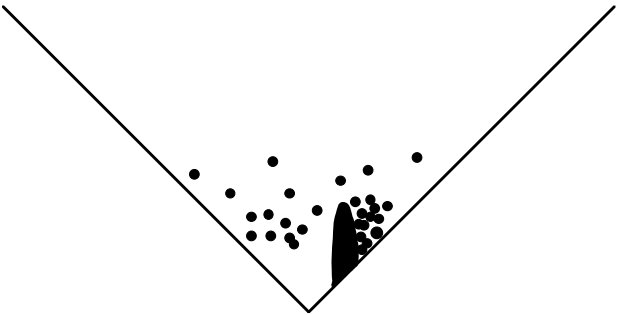
\end{document}